\documentclass[10pt,letterpaper]{article}
\usepackage{opex3,amsmath,braket}
\usepackage{cite}
\usepackage{subfig,hyperref}
\begin{document}

\title{Detuning-Dependent Mollow Triplet of a Coherently-Driven Single Quantum Dot}

\author{Ata Ulhaq,$^{1,2}$ Stefanie Weiler,$^{1}$ Sven Marcus Ulrich,$^{1}$  Michael
Jetter,$^{1}$ and Peter Michler$^{1}$}

\address{$^1$ Institut f\"ur Halbleiteroptik und Funktionelle
Grenzfl\"achen, Allmandring 3, 70569 Stuttgart, Germany\\
$^2$ Department of Physics School of Science and Engineering,
Lahore University of Management Sciences Sector U, DHA, Lahore
54792 Pakistan\\
\color{blue} \textit{\underline{s.weiler@ihfg.uni-stuttgart.de}}}

\author{Chiranjeeb Roy and Stephen Hughes}

\address{Department of Physics, Engineering Physics and Astronomy, Queens's University, Kingston, Ontario, Canada K7L 3N6
\color{blue}\\
\textit{\underline{shughes@physics.queensu.ca}}}


\begin{abstract}
We present both experimental and theoretical investigations of
a laser-driven quantum dot (QD) in the dressed-state regime of resonance fluorescence.
We explore the role of phonon scattering and pure dephasing on
the detuning-dependence of the Mollow triplet and
show that the triplet sidebands may spectrally broaden or
narrow with increasing detuning. Based on a polaron master
equation approach which includes electron-phonon interaction
nonperturbatively, we derive a fully analytical expression for the spectrum.
With respect to detuning dependence, we  identify a crossover between
the regimes of spectral sideband narrowing or broadening.
A comparison of the theoretical predictions to detailed experimental studies
on the laser detuning-dependence of Mollow triplet resonance emission from
single In(Ga)As QDs reveals excellent agreement.
\end{abstract}



\section{Introduction}

Resonant excitation of single QDs has recently
gained a lot of interest
\cite{Muller.Flagg:2007,Flagg.Muller:2009,Vamivakas.Atature:2009}, in part because
this type of coherent excitation is promising for the generation of
single photons with excellent coherence properties
\cite{Kiraz.Atature:2004}. The techniques developed for effective
laser stray light suppression have enabled the collection of
resonance fluorescence from a single QD with high
signal-to-noise ratio. Resonance fluorescence (RF) emission below
saturation of the quantum emitter has revealed close-to-Fourier
limited single photons with record-high emission coherence and
two-photon interference visibility \cite{Ates.Ulrich:2009}. Recent
experiments have even been able to beat the Fourier limit for
single-photon emission coherence in the Heitler regime, e.g., with
excitation strengths well below those to saturate the quantum emitter
\cite{Matthiesen.Vamivakas:2012,Nguyen.Sallen:2011}. Another major
achievement with respect to RF is the demonstration of single- and
cascaded photon emission between the Mollow sidebands above
saturation of the QD \cite{Ulhaq.Weiler:2012}.

Recent investigations of single QD RF have revealed distinct
differences of their emission coherence properties
\cite{Roy.HughesPRX:2011} which need to be theoretically treated
beyond a simple two-level description usually used for atoms.

One of the main consequences of the solid-state character of these quantum emitters
is the consideration of specific dephasing channels primary caused by
carrier-phonon scattering. Dephasing of a resonantly driven QD system
has been theoretically studied in detail with respect to electron-phonon
interaction on the dynamics of an optically driven system
\cite{Foerstner.Weber:2003,Machnikowski.Jacak:2004,AhnFoerstner:2005}.
These studies anticipated excitation-induced dephasing (EID) for
moderate Rabi frequencies. However, non-monotonic behavior was
predicted for Rabi frequencies larger than a cut-off frequency
defined by the material parameters and the QD size
\cite{Vagov.Axt:2007,Nazir:2008}. Non-monotonic
behavior is also predicted for cavity structures with suitable
cavity coupling~\cite{Roy.Hughes:2011,withWaks2011}. Experimental
evidence of EID effects has recently been observed as oscillation
damping in pulsed photocurrent measurements on a resonantly driven
QD \cite{Ramsay.Skolnick:2010}. This damping was found to have a
clear quadratic dependence on the effective Rabi frequency $\Omega$.
The effect of EID has also been observed under strictly resonant continuous wave excitation
of a QD in a microcavity in terms of Mollow-triplet sideband broadening
\cite{Ulrich.Ates:2011}. These experiments reveal good agreement
with a theoretical description based on a polaron master equation
approach to multi-phonon and multi-photon effects in a cavity-QED system
\cite{Roy.Hughes:2011}. In the work of Ulrich {\em et al.}~\cite{Ulrich.Ates:2011},
the phenomenon of spectral Mollow sideband \textit{narrowing} in dependence of
laser-excitation detuning from the bare emitter resonance had to be
left open for further in-depth theoretical analysis.

Motivated by these findings and with the aim of fundamental interpretation,
the focus of the current work lies on a detailed study of the detuning-dependent
dressed state emission of a single QD without cavity coupling. Our
theory is based on a polaron master equation approach from which
we develop a fully analytical description of the emission spectrum.
We are able to distinguish between different regimes of spectral
broadening or narrowing of the Mollow sidebands, under strong
influence by pure dephasing and phonon-induced scattering.
The comparison of detuning-dependent resonance fluorescence data reveals very good
agreement with the theoretical model.

\section{Sample Structure and Experimental Procedure}
\label{sec2} The planar sample employed for the
measurements in this work is grown by metal-organic
vapor epitaxy (MOVPE). The self-assembled In(Ga)As QDs are
embedded in a GaAs $\lambda$-cavity, sandwiched between 29 (4)
periods of $\lambda/4$-thick AlAs/GaAs layers as the bottom (top) distributed Bragg reflectors (DBRs).
For our experimental investigations, the sample is kept in a Helium flow
cryostat providing high temperature stability $T = 5 \pm 0.5$~K. Suppression
of parasitic laser stray-light is achieved by use of an
orthogonal geometry between QD excitation and emission detection.
In addition, polarization suppression and spatial filtering via a pinhole is applied in
the detection path. Resonant (tunable) QD excitation is achieved by a narrow-band ($\approx 500$~kHz)
continuous-wave (cw) Ti:Sapphire ring laser. For high-resolution spectroscopy (HRPL)
of micro-photoluminescence ($\mu$-PL) we employ a scanning Fabry-P\'{e}rot
interferometer with $\Delta E^{\rm HRPL}_{\rm res}< 1\,\mu$eV as described earlier \cite{Ates.Ulrich:2009,Ulhaq.Weiler:2012,Ulrich.Ates:2011}).

\section{Experimental Results: Detuning-Dependent Resonance Fluorescence}
\label{sec3}

\begin{figure}[!b]
\centering
\includegraphics[width= \textwidth]{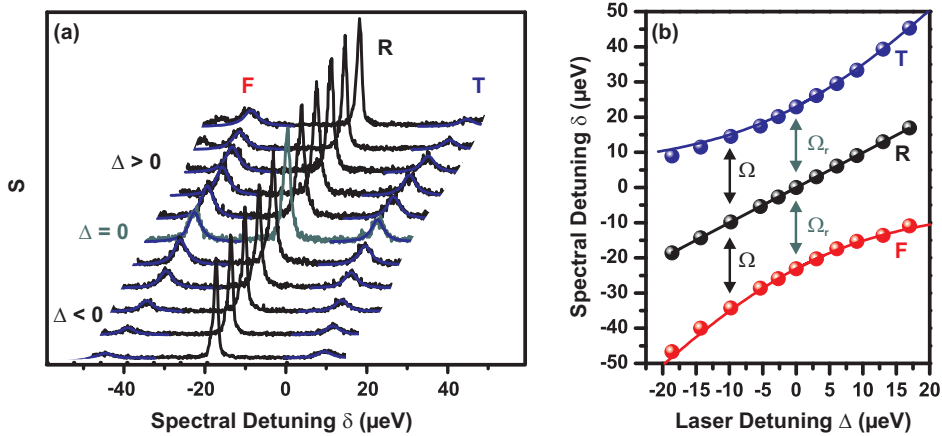}
\vspace{-1.0cm}\caption{(a) HRPL of the QD resonance fluorescence under
systematic variation of the laser-excitation detuning $\Delta =
\omega_L - \omega_x$, taken at a fixed power of $P_0 = 500~\mu$W.
$\delta$ denotes the spectral emission detuning from the bare emitter
resonance. Green center trace: Mollow triplet under strictly resonant
excitation $\Delta = 0$. (b) Spectral evolution of the Mollow
sidebands with laser-detuning $\Delta$, extracted from (a).}
\end{figure}

In our experiments we apply pump powers well above the saturation of
the quantum emitters. In this high-field regime, the excitation-induced Rabi rotation
of the two-level emitter system becomes much faster than the spontaneous decay
rate $\gamma$. The incoherent spectrum of the
resulting dressed state is the characteristic
Mollow triplet \cite{Mollow:1969}. Under strictly resonant
excitation, i.e. for a laser detuning $\Delta = \omega_L - \omega_x = 0$ from the
QD exciton resonance (see Fig.~1(a), green center trace), the spectrum is
composed of the central \textit{Rayleigh line} ``R'' at the bare emitter energy
$\omega_0$ and two symmetric satellite peaks, i.e. the \textit{Three
Photon Line} ``T'' and the \textit{Fluorescence Line} ``F\,'' at $\omega_0 \pm \Omega_r$, respectively.
$\Omega_r$ denotes the effective Rabi frequency including
renormalization effects from the phonon bath as discussed in the
theory section below.

Laser detuning ($\Delta$)-dependent Mollow triplet spectra taken at a
constant excitation strength of $P_0 = 500 ~\mu$ ($\Omega_r \propto (P_0)^{1/2} =$~const.)
are depicted in Fig.~1(a). According to theory
(see, e.g. Ref.~\cite{Vamivakas.Atature:2009}), the laser-detuning between
the driving field and the bare emitter resonance $\omega_0$
modifies the dressed emission. Besides the center transition at
$\omega_0 + \Delta$ the two sideband frequencies become $\omega_0
+ \Delta \pm \Omega$, where $\Omega = \sqrt{\Omega_r^2 + \Delta
^2}$ denotes the generalized Rabi frequency at a given excitation
strength. The extracted spectral positions for the red- and blue-shifted
Mollow sidebands and the central Rayleigh line are depicted in
Fig.~1(b) with a corresponding fit to the data. In addition,
observations on the detuning-dependent Mollow triplet series
depicted in Fig.~1(a) reveal distinct broadening of the Mollow
sidebands with increasing $\Delta$, accompanied by a change in the
relative sideband intensities. In order to explain these
observations, we will develop a theoretical description in
terms of a polaron master equation formalism and derive an analytical
expression for the incoherent spectrum. The main theoretical
findings used to interpret our experimental observations are
described in the following section.

\section{Theory}
\label{sec4}

\subsection{Hamiltonian, Polaron Master Equations and Phonon Scattering Rates}

We model the QD as a effective two-level system interacting
with a coherent pump field and an acoustic phonon reservoir.
In a frame rotating with respect to the laser pump frequency $\omega_L$,
the model Hamiltonian (excluding QD zero-phonon line decay) is
\begin{align}
\label{Ham}
H=-\hbar\Delta\sigma^{+}\sigma^{-} +\hbar \eta_x(\sigma^{+}+\sigma^{-})  +\sigma^{+}\sigma^{-}\sum_{q}\hbar\lambda_{q}(b_{q}
+b_{q}^{\dagger})+\sum_{q}\hbar\omega_{q}b_{q}^{\dagger}b_{q}\, ,
\end{align}

\noindent where $ b_{q}( b_{q}^{\dagger})$ are the annihilation and
creation operators of the phonon reservoir, $\sigma^{+/-}$ (and
$\sigma^z=\sigma^+\sigma^--\sigma^-\sigma^+$) are the Pauli
operators of the exciton. $\eta_x$ is the exciton pump rate, and
$\lambda_q$ (assumed real) is the coupling strength of the
electron-phonon interaction. In order to include electron-phonon
scattering nonperturbatively, we  transform the above
Hamiltonian to the polaron frame. Consequently, we derive a
polaron master equation (ME)~\cite{nazir2,imamoglu,Roy.Hughes:2011,Roy.HughesPRX:2011} which
is particularly well suited for studying quantum optical phenomena
such as resonance fluorescence spectra. In the following we will closely
follow (and extend where necessary) the theoretical formalisms
described in Refs.~\cite{Roy.Hughes:2011,Roy.HughesPRX:2011}, except we can neglect  cavity terms.

Defining $P= \sigma^+ \sigma^-\sum_q \frac{\lambda_q}{\omega_q}
(b_q^\dagger- b_q)$, then the polaron transformed Hamiltonian,
$H^{\prime} \rightarrow e^{P} H e^{-P}$~\cite{mahan}, consists of
a {system} part, {reservoir} part, and an {interaction} part,
respectively:

\begin{align}
\label{sec3eq1ba1}
H^{\prime}_{S} = \hbar(-\Delta-\Delta_{P}){\sigma}^{+}{\sigma}^{-}
+\braket{B} {X}_{g},  \ \ \ \ \ \
H^{\prime}_{R} = \sum_{q}\hbar\omega_{q}{b}_{q}^{\dagger}{b}_{q}, \ \ \ \ \ \ \ \
H^{\prime}_{I} & = {X}_{g}{\zeta}_{g}+{X}_{u}{\zeta}_{u},
\end{align}

\noindent with the coherent displacement operators $B_{\pm}$ defined as
$B_{\pm}=\exp [\pm\sum_{q}\frac{\lambda_{q}}{\omega_{q}}( b_{q}- b_{q}^{\dagger}) ],$ and
${\zeta}_{g}=\frac{1}{2}( B_{+}+ B_{-}-2\braket{B})$ and
${\zeta}_{u}=\frac{1}{2i}( B_{+}- B_{-})$.
The polaron shift is $\Delta_P=\int^{\infty}_{0}d\omega\frac{J(\omega)}{\omega},$
where $J(\omega)=\alpha_{p}\,\omega^{3}\exp (-\frac{\omega^{2}}{2\omega_{b}^{2}})$
denotes the characteristic phonon spectral function that describes
the LA-phonon interaction resulting from deformation potential coupling.
The thermally-averaged bath displacement operator is defined through~\cite{mahan}
$\braket{B}=\exp\left [ -\frac{1}{2}\int^{\infty}_{0}d\omega\frac{J(\omega)}{\omega^{2}}\coth(\beta\hbar\omega/2) \right ]$,
with $\braket{B}=\braket{{B}_{+}}=\braket{{B}_{-}}$, at a bath temperature $T=1/k_b \beta$.
For convenience, we will assume that the polaron shift is implicitly included
in our definition of $\omega_{x}$ below. The  operators $ X_{g}$ and
$X_{u}$ are defined through ${X}_{g} = \hbar \eta_{x}(\sigma^{-}+\sigma^{+})$ and
${X}_{u} = i\hbar \eta_{x}(\sigma^{+}-\sigma^{-}).$

We next present the time-local (or time-convolutionless) ME for
the reduced density operator $\rho(t)$ of the QD-bath system
in the second-order Born approximation of the system-reservoir coupling.
In the interaction picture, we consider the exciton-photon-phonon
coupling $H_{I}^{\prime}$ in the Born approximation and trace over
the phonon degrees of freedom. The \textit{full polaron ME} takes the following form
\cite{nazir2,imamoglu,Roy.Hughes:2011,Roy.HughesPRX:2011}:
\begin{align}
\frac{\partial \rho}{\partial t}&=\frac{1}{i\hbar}[H_{S}^{\prime},\rho(t)]
+\frac{{\gamma}}{2}{\cal L}[\sigma^-]
+ \frac{{\gamma'}}{2}{\cal L}[\sigma_{11}]
\nonumber \\
&-\frac{1}{\hbar^{2}}\int^{t}_{0}d\tau\sum_{m=g,u}
\bigg (G_{m}(\tau)
\left [{X}_{m},e^{-iH_{S}^{\prime}\tau/\hbar}{X}_{m}e^{iH_{S}^{\prime}\tau/\hbar}\rho(t)\right]
+{\rm H.c.} \bigg),
\label{ME1}
\end{align}

\noindent where $\sigma_{11}=\sigma^+\sigma^-$ and the time-dependent function
$G_\alpha(t)\equiv\braket{\zeta_\alpha(t) \zeta_\alpha(0)}$ is derived
as follows: $G_{g}(t)=\braket{B}^{2}\left (\cosh[\phi(t)]-1 \right )$
and $G_{u}(t)=\braket{B}^{2}\sinh[\phi(t)]$~\cite{mahan,imamoglu},
with the phonon correlation function $\phi(t)=\int^{\infty}_{0}d\omega\frac{J(\omega)}{\omega^{2}}
[\coth(\beta\hbar\omega/2)\cos(\omega t)-i\sin(\omega t)]$.
The Lindblad operators ${\cal L}[O]=2O\rho O^\dagger -O^\dagger O\rho - \rho O^\dagger O$
describe dissipation through zero-phonon line (ZPL) radiative decay ($\gamma$) and
ZPL pure dephasing ($\gamma'$), where the latter process is known to increase as a function of temperature~\cite{besombes,BorriPRL:2001,Zimmermann:PRL04,zpl1,12}.
Without the coherent pump term and the residual ZPL broadenings, this ME formally recovers
the independent boson model~\cite{mahan,krum,knorr}.

For continuous wave (cw) excitation, the integration appearing in
Eq.~(\ref{ME1}) can have the upper time limit $t\rightarrow
\infty$, resulting in a Markovian ME where the scattering rates
are computed as a function of $H_S'$~\cite{roy_hughes:PRB2012}.
Such an approach is valid since the acoustic phonon lifetimes are
very fast, i.e. on a few ps timescale. As was shown earlier~\cite{Roy.HughesPRX:2011}, for the pump strengths we consider,
one can neglect the pump-dependence of $H_S'$ appearing in
the exponential phase terms above (which we will further justify below)
to derive an \textit{effective phonon ME} as
\begin{align}
\frac{\partial \rho}{\partial t}&  =\frac{1}{i\hbar}[H_{S}',\rho(t)]+ \frac{\gamma}{2}{\cal L}[\sigma^-] +\frac{\gamma'}{2}{\cal L}[\sigma_{11}]+\frac{\Gamma_{\rm ph}^{\sigma^+}}{2}{\cal L}[\sigma^+] +\frac{\Gamma_{\rm ph}^{\sigma^-}}{2}{\cal L}[\sigma^-] \nonumber \\ & - {{\Gamma{\rm}^{\rm cd}_{\rm ph}}}(\sigma^+\rho\sigma^++\sigma^-\rho\sigma^-) \quad .
\label{ME2}
\end{align}
\noindent Here the pump-driven incoherent scattering processes, mediated by the phonon bath, are
obtained from
\begin{align}
\Gamma^{\sigma^+ / \sigma^-} _{\rm ph}=
\frac{\Omega_r^2}{2} \,{\rm Re} \left [ \int_{0}^\infty d\tau \, e^{\pm
i\Delta\tau} \left (e^{\phi(\tau)}-1\right ) \right ], \nonumber\\
\label{Rates}
\end{align}
and
\begin{align}
\Gamma^{{\rm cd}} _{\rm ph}=
\frac{\Omega_r^2}{2} \,{\rm Re} \left [ \int_{0}^\infty d\tau\,
\cos(\Delta\tau) \left (1-e^{-\phi(\tau)}\right) \right ] \quad .
\end{align}
\noindent The classical Rabi frequency of the exciton pump, including renormalization effects
from the phonon bath, is given by $\Omega_r = {2\eta_x}{\braket{B}}$ (in contrast to
the bare Rabi frequency $\Omega_0={2\eta_x}$).
\begin{figure}[!t]
\centering
\includegraphics[width= 0.62\textwidth]{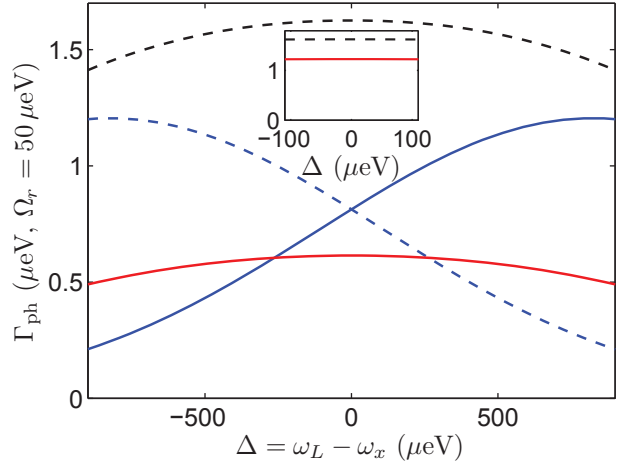}
\vspace{-0.3cm}\caption {\label{thFig1}
Phonon-mediated scattering rates $\Gamma_{\rm ph}^{\sigma^{+/-}}$ (blue solid and dashed lines, respectively), together with their average sum (black dashed line) and $\Gamma_{\rm ph}^{\rm cd}$ (red solid line) for a phonon bath temperature $T = 6$~K. Only for large
detunings $\Delta = \omega_L - \omega_x \geq 0.5~$meV these rates change appreciably. The overall magnitude of the phonon scattering rates is proportional to $\Omega_r^2$. Here we assume $\Omega_r = 50~\mu$eV. As is shown in the inset figure, the sum of rates $\Gamma_{\rm ph}^{\sigma^{+}}+\Gamma_{\rm ph}^{\sigma^{-}}$ as well as $\Gamma_{\rm ph}^{\rm cd}$ are constant within laser-exciton detunings over hundreds of $\mu$eV.}
\end{figure}
The scattering term
$\Gamma_{\rm ph}^{\rm cd}$ is a cross-dephasing rate that
only affects the off-diagonal components of the resulting optical
Bloch equations. Similar terms appear when a system is driven by a
broadband squeezed light reservoir~\cite{Agarwal:1990} and are
sometimes referred to as ``anomalous correlations''. As might be
expected, the excitation-dependent rates depend upon the
phonon correlation function, the coherent pump rate, and the
laser-exciton detuning. The $\Gamma^{\sigma^{-}}_{\rm ph}$ process
corresponds to an \textit{enhanced radiative decay}, while the
$\Gamma^{\sigma^{+}}_{\rm ph}$ process represents an
\textit{incoherent excitation} process~\cite{Roy.Hughes:2011}.
We stress that these mechanisms are quite different to simple pure dephasing models,
which are frequently used to describe weak (i.e., perturbative)
electron-phonon scattering~\cite{Mork:PRL12}. Note that $\Omega_r$ can
be significantly smaller than $\Omega_0$, even at low temperatures.
For example, using InAs QD parameters that closely represent our
experimental samples~\cite{InAsparameters} and a phonon bath temperature $T~6$K,
then $\braket{B} \approx 0.75$, and this value decreases (increases) with
increasing (decreasing) temperature.

For a pump field strength of $\Omega_r = 50~\mu$eV, example phonon
scattering rates are shown in Fig.~\ref{thFig1}. Within the zoomed region of
laser detunings $|\Delta| < 100~\mu$eV the relevant rates
can be assumed to be constant. Therefore, these values will be treated as
constant in the following to compute the analytical Mollow triplets.

\subsection{Mollow Triplet Simulations: Full Polaron versus Effective Phonon ME}

\begin{figure}[th]
\centering
\includegraphics[width= 0.9\textwidth]{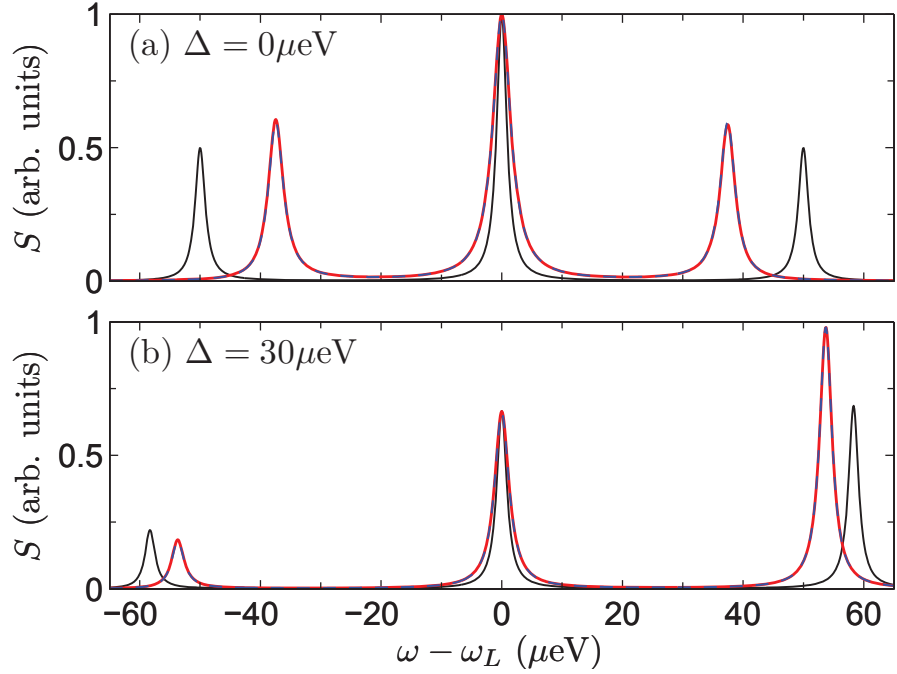}
\caption {\label{thFig2}
 Numerically calculated Mollow triplet with $\Omega_0=50~\mu$eV ($\gamma=\gamma'=1~\mu$eV,
 phonon bath temperature $T=6~$K), for (a) $\Delta=0~\mu$eV and (b) $\Delta=30~\mu$eV,
 plotted for the case of no phonon scattering (black solid
 line), and with phonon scattering using the full polaron ME (red solid line) and the effective ME (blue dashed
 line). The latter two are found to give almost identical
 spectra, which justifies the accuracy of the simpler effective ME.
 Additionally, the effect of renormalization of the Rabi frequency can be
 seen in the different Mollow triplet center-to-sideband
 splittings for the case of no phonon scattering (black line)
 in comparison to the red spectrum.}
\end{figure}

To get better insight into the underlying physics, it is desirable
to derive an analytical form for the Mollow triplet spectrum,
which we derive from the effective phonon ME. Thus we will first
investigate how good the approximation is to replace the phonon
scattering terms in the \textit{full} polaron ME [Eq.~(\ref{ME1})]
by the ones appearing in the \textit{effective} phonon ME
[Eq.~(\ref{ME2})]. In Fig.~\ref{thFig2}, a direct comparison
between the numerically calculated Mollow triplet based on the
full polaron and the effective phonon ME is shown, revealing
excellent agreement even for large detunings $\Delta = 30\,\mu$eV
and high field strengths of $\Omega_0=50~\mu$eV. The main reason
that one can neglect the pump-dependence of the phase terms in
Eq.~(\ref{ME1}) is that---for the pump values we consider---one ps
timescale, phonon correlation times are much faster than the
inverse Rabi oscillation.

We highlight that a cw Rabi field of $\Omega_{r}=50~\mu$eV is
already close to  the highest achievable experiments to date, and
thus for our purposes can be considered the high-field regime.
However, we note that the polaron approach, although
nonperturbative, can break down if extremely high field strengths
are used such that $\Omega_r$ becomes comparable to (or greater
than) the phonon cut-off frequency. In this case, other approaches
exists such as a variational master equation
approach~\cite{VariationalME} and path integral
techniques~\cite{PathIntegral}. Since out maximum Rabi field
strengths are much less than $\omega_b$, as shown by McCutcheon
{\em et al.}~\cite{VariationalME}, the polaron ME should be
rigorously valid for the field strengths that we model.

\subsection{Optical Bloch Equations and Analytical Fluorescence Spectrum}

One of our main theoretical goals is to derive a useful analytical
expression that will allow one to fit the experimental Mollow triplets
over a wide range of parameters, including the laser-exciton detuning.
From the effective phonon ME [Eq.~(\ref{ME2})] and
$\braket{\dot O}={\rm tr}\{\dot \rho O\}$~\cite{Car1},
we obtain the following optical Bloch equations:
\begin{subequations}
  \begin{align}
  \frac{d \braket{\sigma^-}}{dt} = &
  -(\gamma_{\rm pol}+i\Delta)\braket{\sigma^-}-\gamma_{\rm cd}\braket{\sigma^+} -i\frac{\Omega_r}{2} \braket{\sigma^z} , \label{eq:m} \\
  \frac{d \braket{\sigma^+}}{dt} = &
    -(\gamma_{\rm pol}-i\Delta)\braket{\sigma^+}-\gamma_{\rm cd}\braket{\sigma^-} +i\frac{\Omega_{r}}{2} \braket{\sigma^z} , \label{eq:p} \\
  \frac{d \braket{\sigma^z}}{dt} = &
  -i\Omega_r \braket{\sigma^-} +i\Omega_{r} \braket{\sigma^+} -\gamma_{\rm pop} \braket{\sigma^z} - \gamma_{\rm pop}', \label{eq:z}
  \end{align}
\end{subequations}
where we define the polarization decay $\gamma_{\rm pol}=\frac{1}{2}(\Gamma^{\sigma^+}_{\rm ph}+\Gamma^{\sigma^-}_{\rm ph}+\gamma+\gamma')$
and the population decay $\gamma_{\rm pop}=(\Gamma^{\sigma^+}_{\rm ph}+\Gamma^{\sigma^-}_{\rm ph}+\gamma)$,
as well as  $\gamma_{\rm pop}'=\gamma_{\rm pop}-2\Gamma_{\rm ph}^{\sigma^+}$.
For notational convenience, we have also defined $\gamma_{\rm cd} \equiv {\Gamma_{\rm ph}^{\rm cd}}$.
The incoherent spectrum can be computed from an integration of the
appropriate two-time correlation function~\cite{Car1}:
\begin{align}
  S({\bf r},\omega) \equiv  F({\bf r}) S(\omega)  = F({\bf r})\,\frac{1}{\pi} {\rm lim}_{t\rightarrow \infty}{\rm Re} \left \{ \int^{\infty}_{0}d\tau
  \braket{{\delta\sigma^+}(t)\delta{\sigma^-}(t+\tau)}  e^{i(\omega-\omega_{\rm L}) \tau}\right \} ,
\end{align}
where $\braket{\delta O}=\braket{O} -O$ and $F({\bf r})$ is a geometrical factor.
The coherent spectrum can be derived in a similar way. By exploiting the quantum
regression theorem and equation set~(\ref{eq:m})-(\ref{eq:z}), it is possible to derive the
spectrum analytically, e.g., using Laplace transform techniques.
We first define the steady-state expectation values
$f(0)\equiv \braket{\delta\sigma^+\delta\sigma^-}_{ss}$,
$g(0)\equiv \braket{\delta\sigma^+\delta\sigma^+}_{ss}$, and
$h(0)\equiv \braket{\delta\sigma^+\delta\sigma^z}_{ss}$,
and keep the explicit laser-exciton detuning dependence in the solution.
Defining $\delta\omega = \omega-\omega_L$, we can obtain the spectrum from
\begin{align}
  S(\omega)=\frac{1}{\pi}{\rm Re}[f(\delta\omega,\Delta)], \label{eq:Sexact}
\end{align}
where
\begin{align}
  f(\delta\omega,\Delta) = \frac{-f(0)\left [(i\delta\omega-\gamma_{\rm pop})+\frac{{\Omega_r^2}/{2}}{i\delta\omega-(\gamma_{\rm pol}-i\Delta)} \right ]
  +\frac{i\Omega_r}{2} \left [ -h(0) + \frac{i g(0)\Omega_{r}}{i\delta\omega-(\gamma_{\rm pol}-i\Delta)} \right ]}{ \left[i\delta\omega-(\gamma_{\rm pol}+i\Delta)\right](i\delta\omega-\gamma_{\rm pop})+\frac{{\Omega_r^2}
  \left[i\delta\omega-\gamma_{\rm pol} - \gamma_{\rm cd}/2\right]}{i\delta\omega-(\gamma_{\rm pol}-i\Delta)}}. \label{lineshape}
\end{align}
The steady-state inversion and polarization components are
\begin{align}
 \braket{\sigma^s}_{ss}&= - \frac{\gamma_{\rm pop}'}{\gamma_{\rm pop} + \frac{{\Omega_{r}^2}(\gamma_{\rm pol}+\gamma_{\rm cd})/2}
 {(\gamma_{\rm pol}^2 + \Delta^2-\gamma_{\rm cd}^{2})} }, \nonumber \\
 \braket{\sigma^+}_{ss} &=\frac{\frac{-i\Omega_{r}}{2}\left [(\gamma_{\rm pol}+i\Delta)+\gamma_{\rm cd}\right ]}{(\gamma_{\rm pol}^2 + \Delta^2-\gamma_{\rm cd}^{2}) }=(\braket{\sigma^-}_{ss})^*,
\end{align}
from which we can obtain the following steady-state values for $f,\,g$, and $h$:
\begin{subequations}
  \begin{align}
    f(0) & = \frac{1}{2}\left (1+\braket{\sigma^z}_{ss}-\braket{\sigma^+}_{ss}\braket{\sigma^-}_{ss}\right ), \\
    g(0) & = - \braket{\sigma^+}^2_{ss} ,\\
    h(0) & = -\braket{\sigma^+}_{ss}(1+\braket{\sigma^z}_{ss}).
  \end{align}
\end{subequations}
These equations are used with Eq.~(\ref{lineshape}) to obtain $S$.
We stress that the resulting spectrum is \textit{exact} within the stated
model assumptions. The full-width at half-maximum (FWHM) of spectral
resonances can be  obtained from Eq.~(\ref{lineshape}. though these are complicated to write down analytically.
However, as we have verified,  one can simply fit
the analytical spectrum to a sum of Lorentzian line shapes (see discussion of Fig.~4) and extract the broadening parameters.

\begin{figure}[t!]
\centering
      \subfloat[][Spectra for different $r$ values.]
                { \includegraphics[width=0.8\columnwidth]{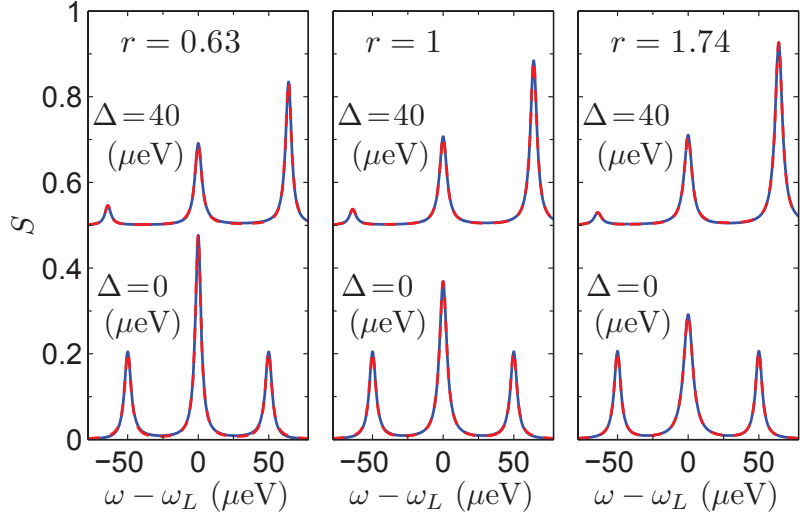}\label{thFig4a} }\\
      \subfloat[][Corresponding sideband FWHM.]
                { \includegraphics[width=0.8\columnwidth]{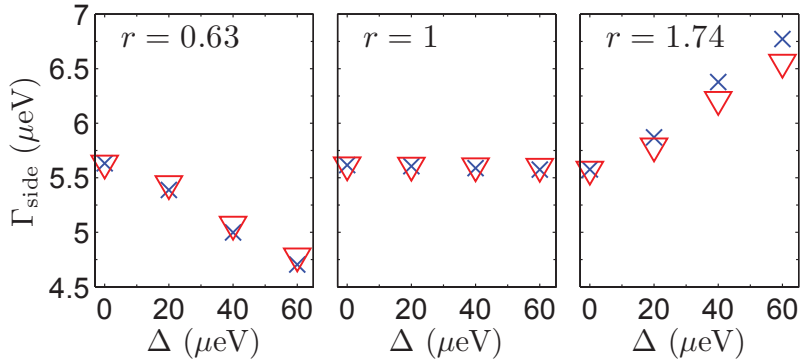}
                \label{thFig4b} }
\caption{(a) Analytically computed spectrum as a function of
detuning for three different values of $r$. The solid red curve
shows the analytical solution [Eq.~(\ref{eq:Sexact})] and the blue
dashed curve shows the three-Lorentzian fit. Positive and negative
detunings $|\Delta|$ reveal simply a mirror image of each other.
The phonon parameters are taken from Fig~\ref{thFig1}, with
$\gamma_{\rm cd} \approx 0.6~\mu$eV and $\gamma_{\rm ph} \approx
1.6~\mu$eV for the chosen Rabi field. Here we adjust $\gamma$ and
$\gamma'$ to maintain the same total low-field FWHM of
$\gamma_{\rm side}(0) = 5.6 ~\mu$eV: For $r = 0.63, 1, 1.74$, we use
$\gamma'(\gamma)$ as $3.2(2.2), 2.5(1.5)$, and $5(1.5)~\mu$eV,
respectively. (b) Extracted FWHM of the lower (blue crosses) and
higher energetic (red inverted triangles) sideband as a function of detuning $\Delta$.
One clearly traces a trend of either increasing or decreasing sideband line width as a
function of laser detuning in dependence on of $r$, where $r \approx 1$ denotes the crossover.}
\end{figure}

\subsection{Off-Resonant Mollow Triplet: Regimes of Spectral Sideband
Broadening and Narrowing}

>From the analytical spectra above, we can discern when the Mollow
triplets will become asymmetric and whether the detuning
dependence will exhibit broadening or narrowing of the three
resonances. We define $r$ as the following ratio:
\begin{align}
  r = \frac{\gamma_{\rm pol}+0.5\,{\gamma_{\rm cd}}}{\gamma_{\rm pop}} = \frac{1}{2}
  \left [1+ \frac{\gamma'+\gamma_{\rm cd}}{\gamma+\gamma_{\rm ph}} \right ],
\end{align}
where $\gamma_{\rm ph} = \Gamma_{\rm ph}^{\sigma^+} + \Gamma_{\rm ph}^{\sigma^-}$.
Worth to note, for off-resonant driving and $\gamma'=0$, a completely
symmetric Mollow triplet is expected \textit{only} if all phonon
terms are neglected. Thus, phonon coupling causes an asymmetry for
off-resonant driving. Under systematic increase of the
excitation-detuning $\Delta$, spectral broadening or narrowing can
be achieved depending upon the value of $r$. In
Figs.~\ref{thFig4a} and \ref{thFig4b} we plot the Mollow triplet
as a function of $\Delta$, and extract the FWHM of
the sidebands for three values of $r$. As can be
seen, $r<1$ (for a suitably small $\gamma'$) leads to
spectral sideband narrowing, whereas the effect of spectral broadening
occurs for values $r>1$.



\section{Comparison between Experiment and Theory}
\label{sec5}
In the following, we show a detailed modeling of experimentally
derived results from detuning-dependent high-resolution PL
measurements with the above presented theory. The measurements are
performed on a QD in a planar sample structure with negligible cavity
coupling.

\begin{figure}[tbp]
\centering
\includegraphics[width= 0.8\textwidth]{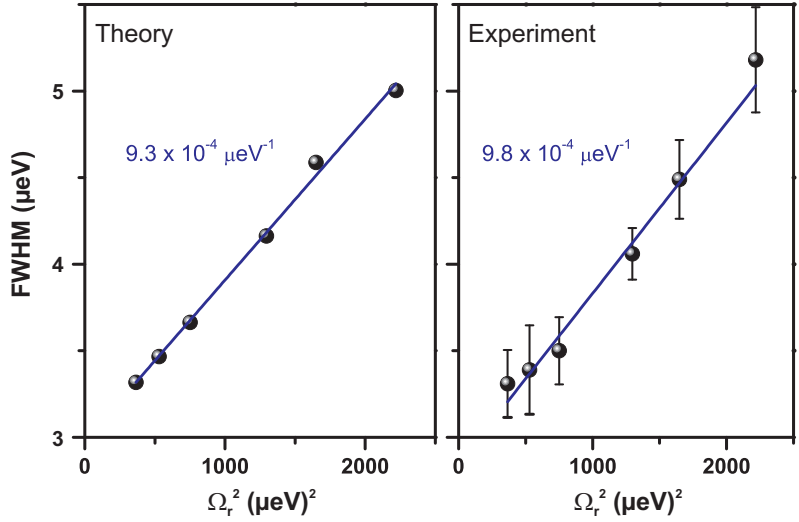}
\caption{Determination of the pure dephasing rate $\gamma'$. The graph shows the
expected linear increase of the FWHM of the Mollow sidebands versus
$\Omega_r^2$ as extracted from a power-dependent Mollow series under
strictly resonant excitation ($\Delta = 0$). A comparison
between (a) the theoretical predictions (revealing a slope of $9.3 \times 10^{-4}
~(\mu {\rm eV})^{-1}$) and (b) the experimental data (giving
$9.8 \times 10^{-4} (\mu {\rm eV})^{-1}$) reveals best consistency for
a dephasing rate of $\gamma' = 4.08\,\gamma = 3.43\,\mu$eV. All other
parameters are fixed according to $T = 6~$K, $\gamma = 0.84 ~\mu$eV (784 ps),
$\alpha_p/(2\pi)^2=0.15~{\rm ps}^2$, $\gamma_{\rm cd}=0.13
~\mu$eV, $\gamma_{\rm ph}=0.34~ \mu$eV. \label{Fig6}}
\end{figure}

To reproduce the experimental results with the theoretical model
we do not use fitting parameters apart from the electron-phonon
coupling strength (which is obtained by fitting the observed EID),
but derive all characteristic values for the calculation via
independent measurements or analysis. For the cut-off frequency
and electron-phonon coupling strength, we use $\omega_b = 1~$meV
and $\alpha_p/(2 \pi)^2 = 0.15~{\rm ps}^2$. The deformation
potential constant is somewhat higher compared to the value used
in Refs.~\cite{Hughes.Yao:2011,Bissiri:2000}. However, the value
for $\alpha_p$ and the dimensionless Huang-Rhys parameter $S_{\rm
HR}= \alpha_p/(2 \pi)^2 c^2_l / l^2_{e/h}$ (with $c_l$ the
speed of sound and $l_{e/h}$  the and electron/hole confinement
length) reported in the literature (i.e. $S_{\rm HR}=0.01-0.5$)
covers a large range and there are no well-accepted numbers to
date. Additionally, $S_{\rm HR}$ has been shown to be enhanced in
zero-dimensional QDs compared to bulk material, for which different
explanations are proposed, e.g., in terms of non-adiabatic effects or
the influence of defects \cite{Bissiri:2000}.

In the experiment the sample temperature has been measured as $T =
6~$K. The QDs in the planar sample are found to have rather
similar radiative lifetimes due to no Purcell-like enhancement.
The radiative decay rate $\gamma$ is extracted from
time-correlated photon counting measurements that reveal a typical
radiative lifetime of (750-850~ps), yielding $\gamma \approx (0.77
- 0.88) \mu$eV. The Rabi field $\Omega_r = 22.7 \mu$eV is derived
from the Mollow center-to-sideband splitting at zero
laser-detuning $\Delta = 0$. The sum of the main phonon scattering
rates $\gamma_{\rm ph} = \Gamma_{\rm ph}^{\sigma^{+}} +
\Gamma_{\rm ph}^{\sigma^{-}} = 0.34~\mu$eV, which is constant in
the detuning range accessible in our measurements, is calculated
according to Fig.~1. The cross-dephasing term has been extracted
from the same graph as $\gamma_{\rm cd}=0.13~\mu$eV. To carefully
extract the pure dephasing rate $\gamma'$, spectra of a
power-dependent Mollow triplet series of the QD under
investigation at $\Delta = 0$ have been modeled with a constant
$\gamma'$ as the only free parameter. The extracted FWHM can be
well reproduced with a pure dephasing rate of $\gamma' =
4.08\,\gamma = 3.43\,\mu$eV (equivalent to a pure dephasing time
of 192~ps). A direct comparison between the extracted FWHM of the
experimental data and the theoretical model is shown in
Fig.~\ref{Fig6}: The expected linear increase [slope: $9.3 \times
10^{-4}~ (\mu {\rm eV})^{-1}$] in the FWHM with $\Omega_r^2$ shows
very good agreement with the experiment [slope: $9.8 \times
10^{-4}~ (\mu {\rm eV})^{-1}$].

\begin{figure}[!b]
\centering
\includegraphics[width=\textwidth]{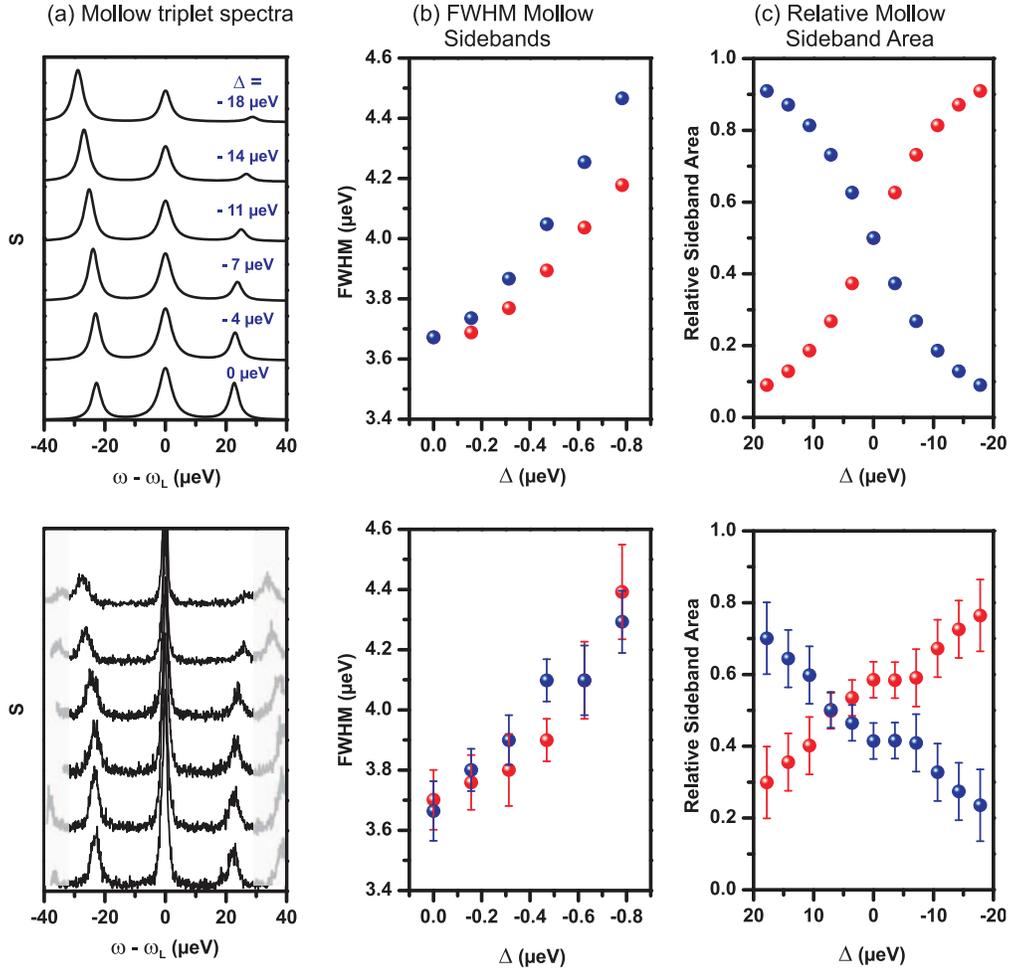}
\caption{Detuning-dependent Mollow triplet series at $P =
500~\mu$eV, showing theoretical predictions versus experimental
results for a system with $r = 2.01$. (a) Mollow triplet spectra
for increasing negative detunings, $\Delta$, the spectra are
plotted with respect to the energetic laser position set to zero.
(b) FWHM of the blue and red Mollow sideband reveal distinct
sideband broadening with increasing laser-detuning. (c) Change of
the relative Mollow sideband area with $\Delta$. The theoretically
expected trend can be seen.}
\label{Fig7}
\end{figure}

With all parameters at hand, the experimentally measured
detuning-dependent Mollow triplet series is compared with the
theoretical expectations in terms of sideband broadening and the
change in the relative sideband areas $A_{\rm red/blue} = I_{\rm
red/blue}/({I_{\rm red}+I_{\rm blue}})$. Figure~\ref{Fig7}(a)
shows a direct comparison of the Mollow triplet spectra for
increasing negative detuning $\Delta < 0$, from which the FWHM and
relative intensities are extracted. The discrepancy between the
expected and measured central Mollow line intensity results from
contributions of scattered laser stray-light to the true QD
emission that can experimentally not be differentiated due to the
equal emission frequency. For the detuning $\Delta \neq 0$ the
spectral resolution of the high-resolution spectroscopy is not
sufficient to distinguish between laser-excitation and QD Rayleigh
line emission. The gray shaded peaks in Fig.~\ref{Fig7}(a) (lower
panel) belong to a higher order interference of the \textit{Fabry
P\'erot} interferometer. The extracted FWHM values are depicted in
Fig.~\ref{Fig7}(b). For the system under investigation, $r$ is
calculated to be around $2.01$, and therefore an increase in the
sidebands' width is expected according to the theoretical model.
Indeed, we observe a systematic increase with increasing negative
detuning $\Delta < 0$. Additionally, the relative sideband areas
$A_{red/blue}$ in dependence on $\Delta$ are plotted in
Fig.~\ref{Fig7}(c). As becomes already visible from the Mollow
spectra, for positive detunings the blue sideband gains intensity
whereas the red sideband area decreases, and vice versa. The
crossing between relative intensities is expected to occur at
$\Delta = 0$. Interestingly, we observe crossings at moderate
negative laser-detuning values for all different QDs under study.
A detailed interpretation of the physics behind this effect has to
be left for further on-going in-depth analysis (and may involve
the inclusion of more excitons). All detuning dependent Mollow
triplet series have revealed values of $r>1$ and therefore a
spectral broadening of the Mollow sidebands due to impurities
present in the sample.

The regime of distinct sideband narrowing has also been
experimentally observed by Ulrich \textit{et~al.}\cite{Ulrich.Ates:2011}
(see Fig.~3(d) of their paper) on a QD embedded in a micropillar
cavity structure, grown by molecular beam epitaxy. Even though the
effect can be qualitatively understood from the theoretical model discussed above,
a direct comparison to theory requires the inclusion of QED-cavity
coupling into the polaron ME approach.
Previous numerical studies were performed in this regime~\cite{roy_hughes:PRB2012}, but did not obtain this behaviour, suggesting that further work is required to explain the significant narrowing effects that are observed in the experiment.

\section{Conclusion}
\label{sec6}
In conclusion, we have presented a combined theoretical-experimental study
on the impact of pure dephasing and phonon-induced scattering on
the excitation detuning-dependence of Mollow triplet sidebands.
Based on a polaron master equation approach, supplemented by an
analytical solution for the Mollow triplet spectra, it is possible to
clearly distinguish different regimes of spectral broadening or narrowing,
defined by the ratio of different dephasing contributions.
For the case of experimentally observed distinct sideband broadening,
we have found excellent agreement with the predictions of theory.
We have derived general formulas which are broadly applicable to reproduce resonance
fluorescence spectra of single quantum dots without adopting a simplistic atomic model.

\section*{Acknowledgements}
During the final preparation of this work we became aware of similar
results obtained independently for off-resonant Mollow sideband
narrowing \cite{McCutcheon.Nazir:2012}. We would like to thank
Dara McCutcheon and Ahsan Nazir for bringing these to our
attention and for useful discussions. S. Weiler acknowledges financial support by the
Carl-Zeiss-Stiftung. The work of S. Hughes  was supported by the
National Sciences and Engineering Research Council of Canada.

\begin{thebibliography}{99}
\bibitem{InAsparameters}
In order to derive the phonon scattering rates, we use parameters
 for InAs/GaAs QDs, which are $\omega_b=1~$meV and
$\alpha_p/(2\pi)^2=0.15~{\rm ps}^2$, where $\omega_{b}$ is the high
frequency cutoff proportional to the inverse of the typical
electronic localization length in the QD
and  $\alpha_p$ is a material parameter (extracted from our  experiments) that
 accounts for the difference between the deformation
potential constants between electrons and holes.
\end{thebibliography}
\end{document}